\documentclass[a4paper,11pt]{article}
\pdfoutput=1 

\usepackage{jcappub} 

\usepackage[T1]{fontenc} 
\usepackage[normalem]{ulem}
\usepackage{amsmath}
\usepackage{amssymb}
\usepackage{geometry} 
\usepackage{booktabs} 
\usepackage{array} 
\usepackage{paralist} 
\usepackage{verbatim} 
\usepackage{graphicx}
\usepackage[utf8]{inputenc}
\usepackage[backend=biber,style=nature]{biblatex}
\usepackage{multirow}
\usepackage[dvipsnames*,svgnames]{xcolor}

\usepackage{wasysym} 
\usepackage{subfigure} 
\usepackage{enumitem}
\usepackage{lineno}

\title{Prospects for the detection of Dark Matter with Long-lived Mediators in the Sun using the Southern Wide-field Gamma-ray Observatory}
\author{Micael Andrade,}
\author{Juan Fagiani,}
\author{Clarissa Siqueira,}
\author{Vitor de Souza,}
\author{Aion Viana}
\affiliation{Instituto de F\'isica de S\~ao Carlos, Universidade de S\~ao Paulo, Av. Trabalhador S\~ao-carlense 400, S\~ao Carlos, Brasil}

\emailAdd{micaelandrade@ifsc.usp.br} 
\emailAdd{juanvitorfagiani@gmail.com}
\emailAdd{csiqueira@ifsc.usp.br}
\emailAdd{vitor@ifsc.usp.br} 
\emailAdd{aion.viana@ifsc.usp.br}

\abstract{The operation of the next generation of gamma-ray observatories will lead to a great advance in dark matter searches. In this paper, we use the hidden sectors hypothesis within the so-called secluded models to calculate the capabilities of the Southern Wide-field Gamma-ray Observatory (SWGO) to detect gamma-ray signatures produced by dark matter particles concentrated in the Sun. We assume the dark matter particle annihilates into metastable mediators which decay into $\gamma\gamma$, $e^+e^-$, $\tau^+\tau^-$, and $\bar{b}b$ outside the Sun. We found that the SWGO will be able to probe a spin-dependent cross-section of about $10^{-46}$~cm$^2$ for dark matter masses smaller than 5~TeV. This result shows an unprecedented sensitivity surpassing the current instruments by more than one order of magnitude. }

\begin{document}

\maketitle
\flushbottom

\section{Introduction}

The most popular hypothesis proposed to solve the Dark Matter~(DM) puzzle is the Weakly Interacting Massive Particle~(WIMP)~\cite{Arcadi:2017kky,Evans:2017kti,Leane:2018kjk,Carena:2019pwq,Bottaro:2021snn,Arcadi:2024ukq}. Searches for WIMPS have been done by experiments based on direct detection via nuclear recoil~\cite{Liu:2022zgu,LZ:2023lvz,XENON:2023cxc,XENON:2024wpa}, by collider experiments through missing transverse momentum~\cite{Abulaiti:2799299,ATLAS:2023ild,Pedro:2023zwd,Slizhevskiy:2023qwj} and by indirect detection experiments, through the stable products of the dark matter annihilation or decay \cite{Fermi-LAT:2016uux,VERITAS:2017tif,CTAConsortium:2017dvg,CTA:2020qlo,Viana:2019ucn,MAGIC:2021mog,LHAASO:2022yxw,HESS:2022ygk,Calore:2022stf,HAWC:2023owv,IceCube:2023ies}. No positive signal of detection for WIMPs has been confirmed so far and the allowed region in the parameter space for several WIMP models is quickly shrinking~\cite{Arcadi:2024ukq}. 

An alternative hypothesis, called secluded models, has been developed~\cite{Pospelov:2007mp,Pospelov:2008jd}. In this scenario, the DM particles interact only indirectly with Standard Model~(SM) particles through dark sector mediators. Secluded models may avoid the strong experimental constraints from direct detection and colliders since the scattering cross-sections of secluded DM particles are very small~\cite{Pospelov:2007mp} when compared to WIMPs. These models allow secluded DM particles to self-annihilate into a dark sector mediator which subsequently decays into SM particles. Given the specific mechanism of production of SM particles from secluded DM particles, the astrophysical signatures of secluded DM particles differ from predictions for WIMPs~\cite{Pospelov:2008jd}. Gamma rays and neutrinos are the final product of the chain of interactions for WIMP or secluded DM particles and therefore they are the smoking guns to detect and differentiate these models. 

An unique signature of secluded DM particles could be produced in many astrophysical objects including the Sun and its planets~\cite{Cermeno:2018qgu,Arina:2017sng,HAWC:2018szf,Niblaeus:2019gjk,Bell:2019pyc,Dasgupta:2019juq,Dasgupta:2020dik,Leane:2021ihh}. DM particles accumulated inside stars and large planets annihilates producing directly (WIMP) or indirectly (secluded) SM particles. In WIMP models, no gamma rays would escape from the Sun and only neutrinos could possibly reach Earth. In secluded models, the SM particles, including gamma rays and neutrinos could be produced outside the Sun reaching Earth without attenuation.

Several observatories searched for gamma-rays~\cite{leane2017powerful,Abeysekara:2013tza,LHAASO:2019qtb} and neutrinos~\cite{Super-Kamiokande:2015xms,ANTARES:2016xuh,IceCube:2021xzo} produced in the Sun. Since no signal has been observed until now, strong constraints have been placed on the spin-dependent scattering cross-section of dark matter. The limits found are orders of magnitude higher than those from direct detection experiments~\cite{leane2017powerful}. 

In this work, we address for the first time the sensitivity of the future Southern Wide-field Gamma-ray Observatory (SWGO) to the dark matter spin-dependent scattering cross-section for secluded models. SWGO is a future water Cherenkov observatory to be constructed in South America. It will be able to detect gamma rays spanning from hundreds of GeV to the PeV scale in energy. It is expected that the SWGO's sensitivity will be one order of magnitude better than the current water Cherenkov gamma-ray experiments~\cite{Albert:2019afb}. One of the primary motivations for establishing this gamma-ray observatory in the Southern Hemisphere is the unique opportunity to observe the Galactic Center and the Galactic plane~\cite{Viana:2019ucn}. Given its non-optical technology, SWGO will also conduct solar science research, including DM searches in the Sun~\cite{Albert:2019afb}. 

In section~\ref{sec:dmsun}, we compute the gamma-ray flux produced in the annihilation of secluded DM particles into metastable mediators inside the Sun. In section~\ref{sec:swgo}, we use the instrument response function of SWGO to simulate the signal to be detected when the observatory is running. In section~\ref{cha:results}, we show the analysis that could lead to detection or a limit on secluded DM features. Section~\ref{sec:conc} concludes our work showing that SWGO will be able to probe a spin-dependent cross-section of about $10^{-46}$~cm$^2$ for dark matter masses smaller than 5~TeV.

\section{Production of gamma-rays by secluded dark matter annihilation in the Sun}
\label{sec:dmsun}

\subsection{Secluded Dark Matter}

One of the most interesting and intriguing ways to escape from the current strong limits of direct detection and collider experiments is by considering that the DM particle's direct interaction with SM particles is highly suppressed~\cite{Pospelov:2007mp}. In these scenarios, we have a dark sector where the DM particles and their mediators lie. They are usually called secluded or hidden sectors or cascade annihilation models. Although this would, in principle, inhibit their collision rate in underground WIMP detectors or their production in colliders, the flux of SM particles that can be created as a sub-product of the DM particle annihilation would remain almost unchanged. Hence, these models provide ideal candidates for indirect DM searches, and several experiments have been looking for possible signals in a vast range of DM particle masses.

The portals connecting the dark sector to the standard sector can be a scalar, a vector, or a right-handed neutrino~\cite{Pospelov:2007mp}. The first secluded DM models were proposed in a low mass range (at the scale of MeV), intending to explain the 511 keV line observed at the galactic center by SPI/INTEGRAL~\cite{Knodlseder:2003sv,Jean:2003ci,Knodlseder:2005yq} and also to avoid the limits from direct and collider searches~\cite{Pospelov:2007mp}. They were also considered as candidates for the GeV galactic center excess, avoiding the gamma-ray limits from dwarf spheroidal galaxies provided by the Fermi-LAT \cite{Martin:2014sxa,Elor:2015tva,Dutta:2015ysa,Escudero:2017yia}. Undoubtedly, secluded models have been overly studied in the literature in several contexts, unveiling an alternative way to look for DM particles.

From a theoretical point of view, secluded scenarios can appear in several ways, including minimal $U(1)$ extensions of the standard model (SM), where the DM is a fermion and the mediator is a vector boson, under the condition that the DM is more massive than the mediator's, and the channel $DM + DM \rightarrow V + V$ is kinematically open~\cite{Fortes:2015qka}. Also, we may have a scalar mediator, with the DM being a singlet fermion, which, under the same condition, provides a secluded scenario~\cite{Pospelov:2007mp}. It is important to emphasize that in both cases the mediator is metastable and decays into standard model pairs, as $l^+ l^-$, $\bar{q}q$, or even $\gamma\gamma$ \cite{Arina:2017sng}. Another possibility is a fermionic mediator, they may appear in the context of seesaw models to explain neutrino masses, in this case, the DM candidate can be a scalar or a fermion~\cite{Cosme:2020mck}.

Several works were done studying the sensitivity of current and future experiments to detect possible signals of secluded models~\cite{Elor:2015bho,Profumo:2017obk,Siqueira:2019wdg,NFortes:2022dkj}, from different targets and possible final states, strongly constraining the velocity averaged annihilation cross-section of these DM particle candidates. Besides the standard search for prompt annihilation emission in DM halos, it is also possible to look for secluded DM particles in stars, planets, and other astrophysical objects. In the absence of detection of a signal coming from these targets, it is possible to put strong limits on the spin-dependent DM cross-section. We will discuss the details in the following sections. 

\subsection{Dark matter capture and annihilation inside the Sun}

It has been shown that the Sun accumulates DM by capturing gravitationally the DM particles in the galactic halo~\cite{sivertsson2010captureprocess}. Considering secluded DM particles in the Sun, its annihilation generates long-lived metastable mediators that can decay outside the Sun. If the decay length ($L$) of the mediator is larger than the radius of the Sun, an attenuated signal can be detected at Earth.

If we ignore self-interaction~\cite{Zentner:2009is}, the number of DM particles $N_\chi(t)$ inside the Sun at a given time $t$ can be simply given by the capture and annihilation rates as
\begin{equation}
    \frac{dN_\chi(t)}{dt} = \Gamma_C - C_A N^2_\chi(t),
\end{equation}
where $\Gamma_C$ corresponds to the capture rate and $C_A$ is a coefficient that encompasses the annihilation cross section and the DM number density. Once equilibrium between capture and annihilation is reached ($dN_\chi(t)/dt = 0$), the annihilation rate will be given by
\begin{equation}
    \Gamma_A = \frac{1}{2} C_A N^2_\chi = \frac{1}{2} \Gamma_C \, ,
\end{equation}
independent of the velocity-averaged annihilation cross-section $\langle \sigma_A v \rangle$. Thus, $\Gamma_A$  will depend mainly on the scattering cross-section and the local halo mass density. Using this result, the expected gamma-ray flux from the DM annihilation inside the Sun is given by
\begin{equation}
    \frac{d\Phi_\gamma}{dE} =  \frac{\Gamma_C}{8\pi D_{sun}^2} \sum_i BR_i \frac{dN_i}{dE} P_{S} \, ,
    \label{eqn:flux}
\end{equation}
where $D_{sun}$ represents the Sun-Earth distance, $BR_i$ the branching ratio, $dN_i/dE$ the annihilation spectrum, with $i$ representing the annihilation channel, and $P_{S}$ is the probability that the mediator will decay outside the Sun and before reaching Earth
\begin{equation}
    P_{S} = e^{-R_{\odot}/L}-e^{-D_{sun}/L}. 
    \label{eqn:pobs}
\end{equation}
where $L$ is the decay length of the mediator and $R_{\odot}$ is the radius of the Sun. In this work, we assume optimal signal conditions given by $L= \gamma c \tau \approx R_{\odot}$ which results in  $P_S \approx 0.4$~\cite{leane2017powerful}.

In this calculation, we neglected the evaporation of DM inside the Sun because its effect is considerable only for DM masses below a few GeV~\cite{griest1987cosmicevap, spergel1985effectevapcap, gould1987weaklyevap}. We assume a branching ratio equal to 100\% for different annihilation channels. This assumption is made to ensure the model's independence from specific theoretical frameworks, making it relatively straightforward to adapt and rescale for various models with specific branching ratios. 

\begin{figure}[h!]
    \begin{center}
        \includegraphics[width=0.9\textwidth]{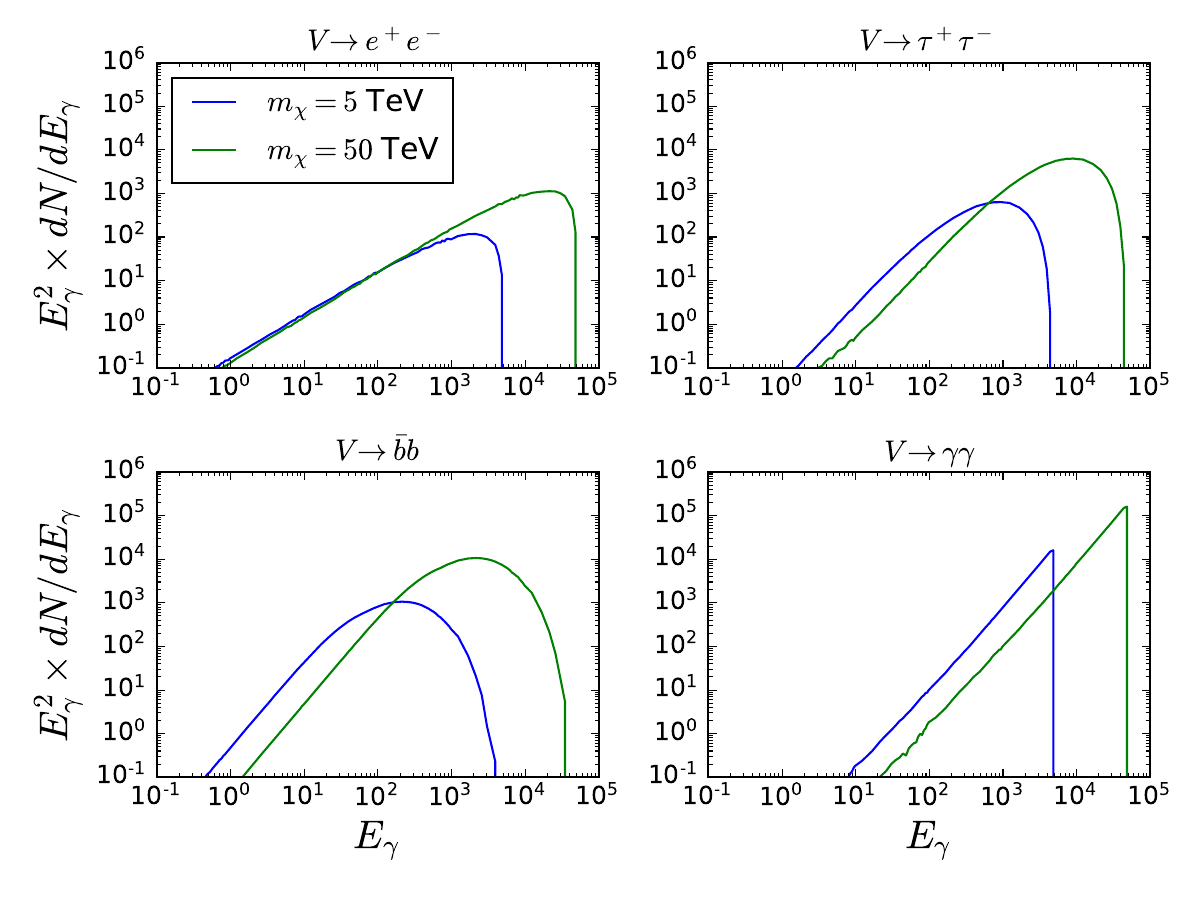}
        \caption{Gamma-ray energy spectra produced by the annihilation of secluded DM matter with mass $m_\chi=5$ and  $50$ TeV for the channels $V \rightarrow\gamma\gamma$, $V\rightarrow e^+e^-$, $V\rightarrow \tau^+\tau^-$, and $V \rightarrow\bar{b}b$. The calculation was done using the Pythia 8.3 package~\cite{Bierlich:2022pfr}. See text for details. }
        \label{fig:spec}
    \end{center}
\end{figure}

The gamma-ray spectrum $dN/dE_\gamma$ for each channel was computed using the software package Pythia 8.3~\cite{Bierlich:2022pfr}, assuming $m_\chi\gg m_{V}$. Figure~\ref{fig:spec} shows the energy spectra of gamma-rays produced in the annihilation of secluded DM with masses of $m_\chi=5~\rm{TeV}$ and  $m_\chi=50~\rm{TeV}$ for four channels:$V \rightarrow\gamma\gamma$, $V\rightarrow e^+e^-$, $V\rightarrow \tau^+\tau^-$, and $V \rightarrow\bar{b}b$.

\subsection{Relation between capture rate and scattering cross-section}

The scattering cross-section between WIMPs and nuclei can be separated into two distinct components: spin-dependent (SD) and spin-independent (SI). In the first case, the cross-section directly hinges on the spin of the target nucleus. In a differential sense, this relationship can be expressed as $d\sigma_{\chi N}^{\mathrm{SD}}/dE \propto J(J+1)$ (being $J$ the spin of the nucleus). Conversely, the second component scales with the square of the target nucleus's mass number represented as $d\sigma_{\chi N}^{\mathrm{SI}}/dE \propto A^2$ \cite{Bertone:2010zza}. Consequently, when it comes to direct detection experiments for $\sigma_{\chi N}^{\mathrm{SI}}$, they exhibit sensitivity that is orders of magnitude greater (approximately $\approx 10^6$) than SD searches. Enhancing this sensitivity is a matter of selecting heavier target atoms.

In contrast, improving sensitivity for $\sigma_{\chi N}^{\mathrm{SD}}$ is significantly constrained by the need to increase the spin. This scenario of probing the Sun for DM with gamma-rays not only enables us to establish more stringent constraints on $\sigma_{\chi N}^{\mathrm{SD}}$ but also serves as a complementary and independent approach to direct detection experiments focused on investigating the nature of DM and its interactions with baryonic matter.

The relation between the spin-dependent scattering cross-section $\sigma^{SD}_{\chi p}$ and the capture rate $\Gamma_C$ can be found through the use of a simplified formula, which works as an estimate for the case when in thermal equilibrium \cite{bertone2005particle,profumo2019introduction,gould1987Tcapture, Rott_2011},
\begin{equation}
    \Gamma_C = 3.4\times10^{20}\, s^{-1} \bigg(\frac{\rho^\odot_\chi}{0.3\,\textrm{GeV/cm}^3}\bigg) \bigg(\frac{270\, \textrm{km/s}}{v_\chi}\bigg)^3 \bigg(\frac{\sigma^{SD}_{\chi p}}{10^{-42}\,\textrm{cm}^2}\bigg) \bigg(\frac{100\,\textrm{GeV}}{m_\chi}\bigg)^2,
    \label{eqn:capture}
\end{equation}
with this relation and the signal found through equation~\ref{eqn:signal}, we are then able to find the sensitivity of gamma-ray telescopes to the spin-dependent DM scattering cross-section between DM and nuclei.

\section{Detector simulation and data analysis}
\label{sec:swgo}

The simulation of the detection of the gamma-ray flux calculated in the previous sections is done using the instrument functions (IRFs) produced by the SWGO Collaboration for a Strawman design of the observatory\footnote{\url{https://github.com/harmscho/SGSOSensitivity}}. The application of the IRFs on the energy spectrum of gamma-rays produced by secluded DM as calculated in the previous sections, results in the energy spectrum of gamma-rays detected by SWGO including experimental uncertainties. 
 
The statistical analysis used to derive the sensitivity of SWGO to the detection of a DM-induced gamma-ray flux from the Sun is the log-likelihood ratio test statistic. The expected count of gamma-ray events (or signal) $S_i$ detectable by a telescope with an effective area of $A_{eff}$ in an observation time period $T$, in an energy interval $\Delta E_i$, is estimated by
\begin{equation}
    S_i = T\int_{\Delta E_i} d \hat{E} \int_0^\infty dE \frac{d\Phi_\gamma}{dE}(E) A_{eff}(E) PDF(E,\hat{E}),
    \label{eqn:signal}
\end{equation}
where $PDF(E,\hat{E})$ represents the energy resolution of the instrument as the probability density function $P(\hat{E}|E)$ of observing an event at the reconstructed energy $\hat{E}$ for a given true energy $E$ \cite{Viana:2019ucn}. Since $S_i$ is a function of $d\Phi_\gamma / dE $, it can also be written as a function of the DM capture rate, and hence the scattering cross-section between DM and nuclei (see Eq.~\ref{eqn:flux} and Eq.~\ref{eqn:capture}).   


The likelihood function $\mathcal{L}$ is defined as~\cite{1983ApJ...272..317L, Viana:2019ucn}
\begin{equation}
    \mathcal{L} = \prod_i \mathcal{L}_i = \prod_i \frac{(S_i+B_i)^{N_i} e^{-(S_i+B_i)}}{N_i!},
\end{equation}
where $S_i$ is the expected signal, $B_i$ the expected background, and $N_i$ the observed counts in each energy bin $i$. In case of no detection ($N_i = B_i$) by SWGO, we can obtain the exclusion limits by comparing an alternative hypothesis $\mathcal{L}_i$ of detecting a positive signal ($S_i >0$) to a null hypothesis ($\mathcal{L}_0$: $S_i=0$), using a test statistic ($TS$) defined as
\begin{equation}
    TS (\Gamma_C) = \sum_i \begin{cases}
    -2\ln(\mathcal{L}_i(\Gamma_C)/\mathcal{L}_0),& \text{if } B_i\geq1\\
    2S_i(\Gamma_C),              & \text{otherwise.}
\end{cases}
\end{equation}
The 95\% confidence level (C.L.) limits are found for values of the capture rate $\Gamma_C$ where $TS = 2.71$, and for each DM particle mass $m_{\chi}$, these can be translated into limits to the spin-dependent scattering cross-section. 

\begin{figure}[h!]
    \begin{center}
        \includegraphics[width=1\textwidth]{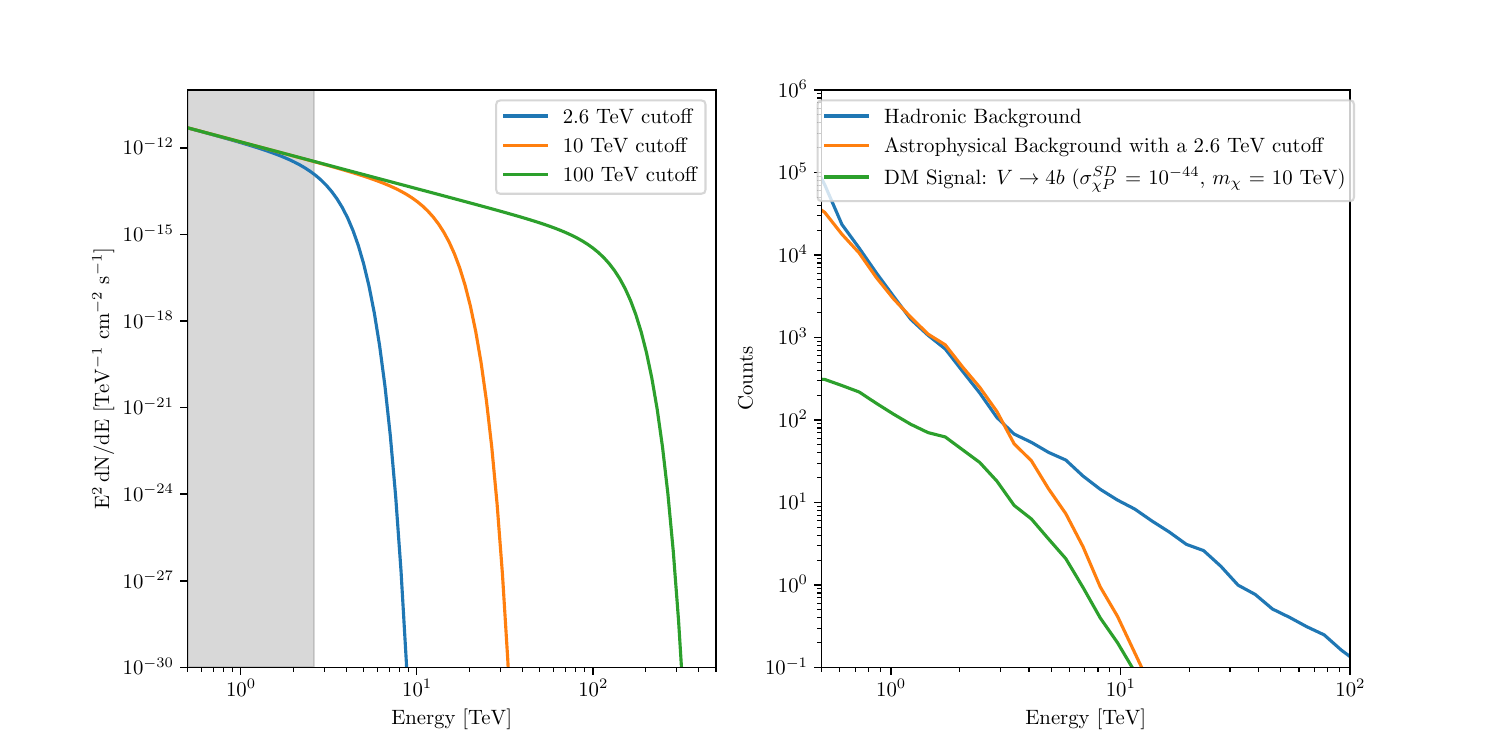}
        \caption{\textit{Left: }The astrophysical spectra, \( dN/dE \), from the Solar disk as measured by HAWC, are presented with different exponential cut-off values, using an index of 3. The grey-shaded region indicates the energy range used by HAWC to determine the spectrum. \textit{Right: }The expected signal for dark matter annihilation inside the Sun via the $V \to 4b$ channel is calculated, assuming 10 years of observation, a scattering cross-section of $\sigma^{SD}_{\chi P} = 10^{-44}$ cm\(^2\), and a dark matter mass of $m_{\chi} = 10$ TeV. For comparison, the expected background counts from both hadronic (pos-cut) and astrophysical components are also plotted.}
        \label{fig:spectra_cutoff}
     \end{center}
\end{figure}

For Sun observations using ground-based water Cherenkov observatories, two background components must be considered. The first is the usual residual hadronic background, originating from misidentified cosmic-ray showers following gamma-hadron separation. This background is provided by the SWGO IRFs, with a 1\% reduction applied due to the Sun-shadow effect~\cite{HAWC:2018rpf}. The second component is the astrophysical gamma-ray background from the Sun's disk, recently reported by the HAWC collaboration \cite{HAWC:2022khj}. The exact mechanism producing this emission is not yet fully understood, but it is believed to be caused by hadronic Galactic cosmic rays interacting with nuclei in the solar atmosphere. The gamma-ray flux was detected in the 0.5–2.6 TeV range, with a spectrum following a power-law with an amplitude of $(1.6 \pm 0.3) \times 10^{-12}$ TeV$^{-1}$ cm$^{-2}$ s$^{-1}$ at 1 TeV and an index of 3.62. To estimate the effect of extrapolating this background into the multi-TeV range, we introduced an exponential cut-off at various energies and analyzed its impact on our sensitivities. Figure \ref{fig:spectra_cutoff}\emph{-left} shows the observed solar disk gamma-ray spectrum with cut-offs at 2.6, 10, and 100 TeV. In Figure~\ref{fig:spectra_cutoff}\emph{-right}, we also compare the expected signal for a specific dark matter annihilation inside the Sun, via the $V \to 4b$ channel, with the expected counts for the two background components considered.

\section{Results and comparison with current limits}
\label{cha:results}

\begin{figure}[h!]
    \begin{center}
        \includegraphics[width=0.9\textwidth]{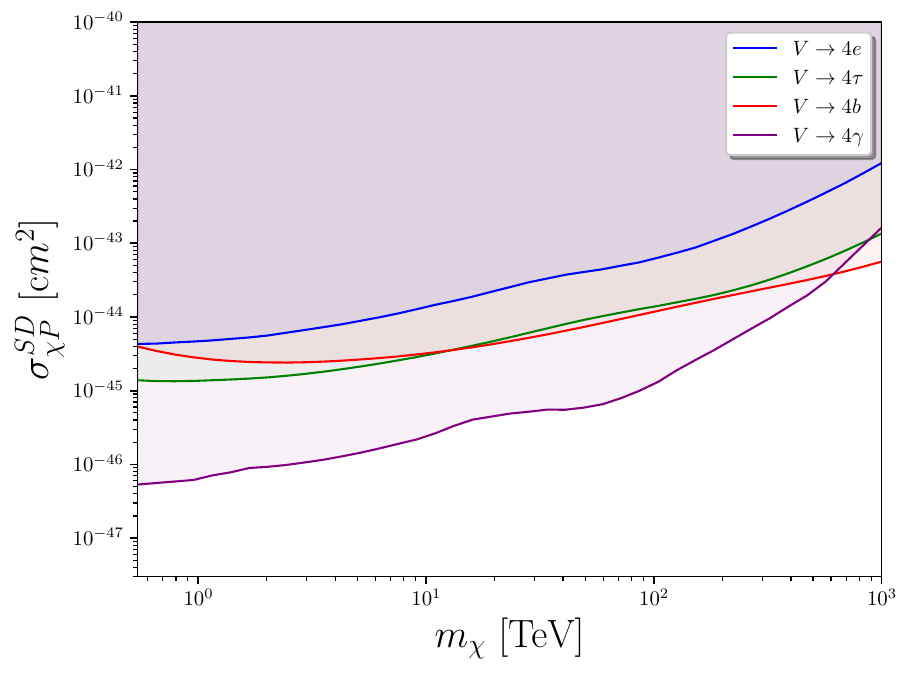}
        \caption{Upper limits at 95\% C.L of the spin-dependent scattering cross-section of DM-nuclei $\sigma^{SD}_{\chi p}$ as a function of the DM particle mass $m_\chi$ for four annihilation channels $b\overline{b}$, $e^+e^-$, $\tau^+\tau^-$, and $\gamma\gamma$. We considered a 100\% branching ratio for each channel.}
        \label{fig:onepanel_limits}
     \end{center}
\end{figure}

The DM-nuclei spin-dependent scattering cross-section $\sigma^{SD}_{\chi p}$ limits are computed for DM annihilation within the Sun, for the Strawman design of the SWGO experiment and assuming a 5-year operation period. Figure~ \ref{fig:onepanel_limits} shows the upper limits derived for four distinct annihilation channels, $b\overline{b}$, $e^+e^-$, $\tau^+\tau^-$, and $\gamma\gamma$, and DM particle masses from 500 GeV to 1 PeV. We only compute results for the SD scenario, as the SI case is already constrained by much stronger limits from direct detection experiments \cite{2023PhRvL.131d1002A}.

\begin{figure}[h!]
    \begin{center}
        \includegraphics[width=1\textwidth]{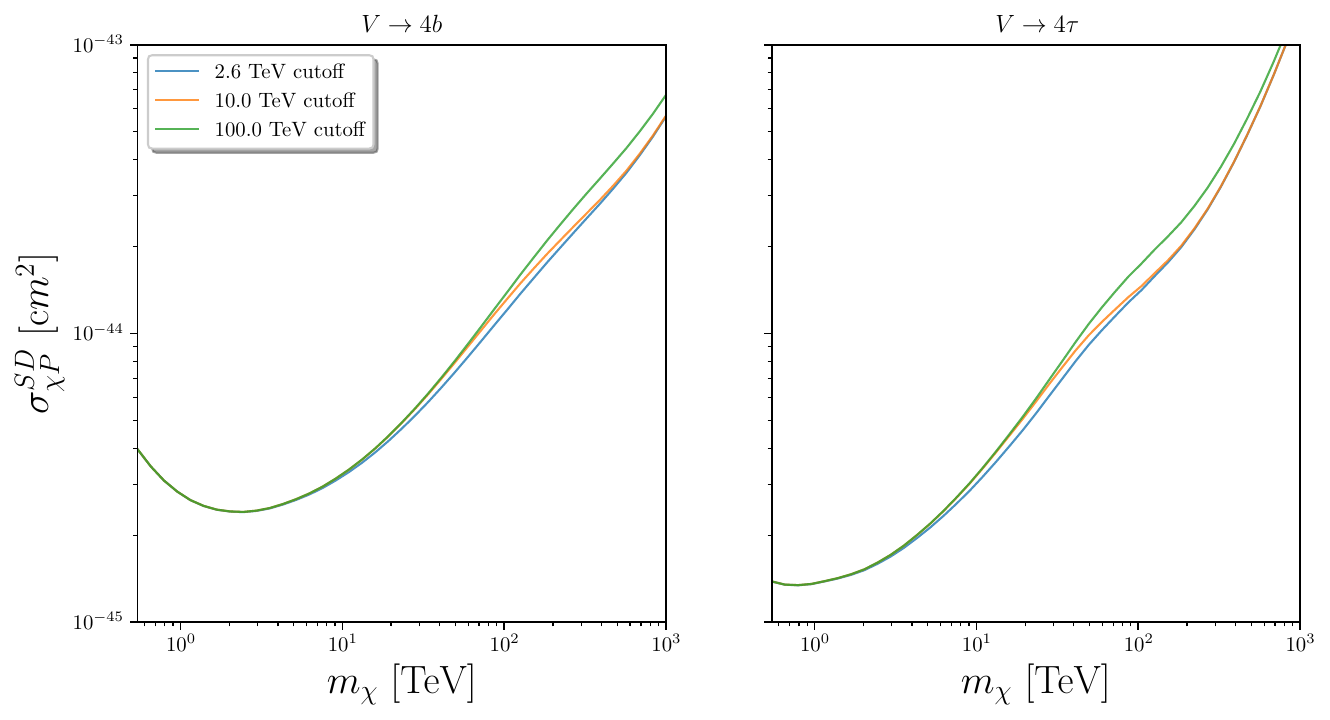}
        \caption{Comparison of the impact of different exponential cut-off values on the astrophysical gamma-ray background from the Sun disk detected by HAWC~\cite{HAWC:2022khj}. The plot shows the 95\% C.L. upper limits on the spin-dependent DM-nuclei scattering cross-section $\sigma^{SD}_{\chi p}$ as a function of the DM particle mass $m\chi$, for three distinct cutoff values. A 100\% branching ratio for the $b\overline{b}$ and $\tau^+\tau^-$ channels was assumed.}
        \label{fig:cutoff_comparison}
     \end{center}
\end{figure}

The channel with the hardest spectra $\gamma\gamma$ has the most constraining results for the entire range of DM particle mass analyzed, as expected due to the larger signal-to-noise ratio in each energy bin. For this channel, the spin-dependent scattering cross-section $\sigma^{SD}_{\chi p}$ reaches $1\times 10^{-46}\mathrm{ cm}^2$ at $m_\chi < 1$ TeV. As $m_\chi$ increases, the advantage of the $\gamma\gamma$ channel becomes less pronounced. At $m_\chi \sim 1$~PeV the upper limit for the $\gamma\gamma$ and the $b\overline{b}$ channels are similar. For all these results, we used a gamma-ray background from the Sun disk with a cut-off value of $2.6$ TeV. However, this choice has a negligible impact on the limits, as shown in Figure~\ref{fig:cutoff_comparison}, which compares the limits using different cut-off values.

\begin{figure}[h!]
    \begin{center}
        \includegraphics[width=0.9\textwidth]{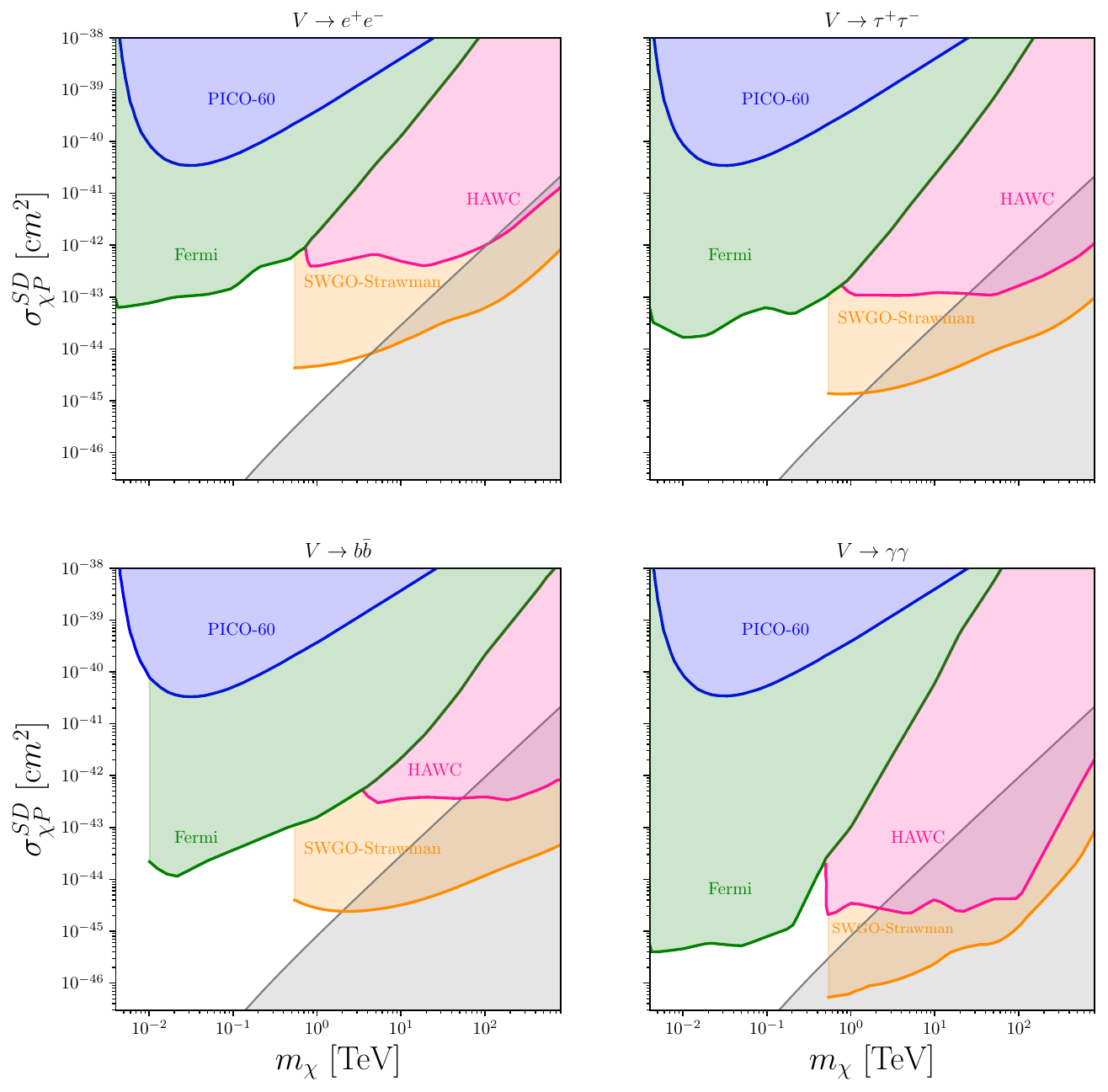}
        \caption{Upper limits at 95\% C.L of the spin-dependent scattering cross-section of DM-nuclei $\sigma^{SD}_{\chi p}$ as a function of the DM particle mass $m_\chi$ for four annihilation channels $b\overline{b}$, $e^+e^-$, $\tau^+\tau^-$, and $\gamma\gamma$. We show the comparison of results obtained here for SWGO with the previous analysis done with the HAWC and Fermi-LAT data~\cite{HAWC:2018szf} and with PICO-60 data~\cite{PICO:2017tgi}. Additionally, we include the thermalization limits for the scattering cross section, taking into account the age of the solar system, shown in grey. We considered a 100\% branching ratio for each channel.}
        \label{fig:comparisonlimits}
    \end{center}
\end{figure}

Figure~\ref{fig:comparisonlimits} presents a comparison between our results and the limits published using HAWC and Fermi-LAT data~\cite{HAWC:2018szf}, as well as the most stringent limits from direct detection experiments provided by PICO60-C$_3$F$_8$~\cite{PICO:2017tgi}. Each of the four annihilation channels is shown in separate panels. Additionally, we include the thermalization limits, which indicate the threshold below which the capture process would not have had sufficient time to reach equilibrium since the Sun's formation ($\approx4.5$ billion years ago) \cite{Peter:2009mk,Peter:2009mi}. For masses and cross-section values below this threshold, the dark matter flux would be severely suppressed, rendering our initial hypothesis invalid.

We show that SWGO will have the potential to further improve the constraints on the spin-dependent scattering cross-section for all DM particle mass above 500 GeV. For DM mass around 1 TeV, the improvement is better than one order of magnitude in comparison to the HAWC and Fermi-LAT data. These results clearly showcase the SWGO's increased capability in the lower range of the energy spectrum when compared to HAWC, an advantage that arises, in particular, from having a larger array with a higher fill factor.

\section{Conclusion}
\label{sec:conc}

Gamma-ray observations of the Sun by water Cherenkov detector provide a new avenue to detect DM particles with long-lived mediators. This capability is particularly interesting as it complements indirect searches for DM done by Imaging Atmospheric Cherenkov telescopes, such as the future CTA Observatory~\cite{CTAConsortium:2017dvg}. 

In this work, we presented SWGO capabilities to detect such DM particles, and in the case of no detection, sensitivity limits to the spin-dependent cross-section of DM-nuclei interactions were derived. It was shown that SWGO will improve current limits by orders of magnitude concerning leading direct detection experiments, and up to two orders of magnitude when compared to indirect detection experiments, such as HAWC and Fermi-LAT. A natural progression of the current study would involve extending these constraints to encompass various decay lifetimes and diverse targets, similar to the approach taken with Jupiter and targets within the Galactic Center \cite{Leane:2021ihh, Leane:2021tjj}. 

However, for a comprehensive and complete analysis, utilizing SWGO final design IRFs is essential. Nevertheless, these findings not only underscore the potential for solar science with SWGO but also emphasize the unique opportunities this type of experiment presents.

\section*{Acknowledgments}

Authors are supported by the S\~{a}o Paulo Research Foundation (FAPESP) through grants number 2019/14893-3, 2020/00320-9, 2021/01089-1, 2023/15494-0. CS is supported by FAPESP through grant number 2020/00320-9. AV and VdS are supported by CNPq through grants number 314955/2021-6 and 308837/2023-1, respectively. The authors acknowledge the National Laboratory for Scientific Computing (LNCC/MCTI,  Brazil) for providing HPC resources for the SDumont supercomputer (http://sdumont.lncc.br).

\printbibliography



\end{document}